# Consumer's Behavior Analysis of Electric Vehicle using Cloud Computing in the State of New York

Jairo Juarez, Wendy Flores, Zhenfei Lu, Mako Hattori, Melissa Hernandez, Safir Larios-Ramirez,
Jongwook Woo
California State University Los Angeles, FEMBA
e-mail : {jjuare151, wflore17, zlu19, mhattor, mhern283, slarios, jwoo5}@calstatela.edu

**Abstract:** Sales of Electric Vehicles (EVs) in the United States have grown fast in the past decade. We analyze the Electric Vehicle Drive Clean Rebate data from the New York State Energy Research and Development Authority (NYSERDA) to understand consumer behavior in EV purchasing and their potential environmental impact. Based on completed rebate applications since 2017, this dataset features the make and model of the EV that consumers purchased, the geographic location of EV consumers, transaction type to obtain the EV, projected environmental impact, and tax incentive issued. This analysis consists of a mapped and calculated statistical data analysis over an established period. Using the SAP Analytics Cloud (SAC), we first import and clean the data to generate statistical snapshots for some primary attributes. Next, different EV options were evaluated based on environmental carbon footprints and rebate amounts. Finally, visualization, geo, and time-series analysis presented further insights and recommendations. This analysis helps the reader to understand consumers' EV buying behavior, such as the change of most popular maker and model over time, acceptance of EVs in different regions in New York State, and funds required to support clean air initiatives. Conclusions from the current study will facilitate the use of renewable energy, reduce reliance on fossil fuels, and accelerate economic growth sustainably, in addition to analyzing the trend of rebate funding size over the years and predicting future funding.

### 1. Introduction

This paper uses SAP Analytics Cloud (SAC) to process and visualize the data of electric vehicle drive clean rebates in New York State. SAC runs as a cluster. The dataset was retrieved from Data.gov and provided by NYSERDA. It mainly consists of information on electric vehicle rebate amounts, type of electric vehicle including make and model, annual greenhouse gas (GHG) emissions reductions, annual petroleum reductions in gallons, whether the vehicle was purchased or leased, across different counties and zip codes of the state of New York as of 2017. We have chosen this dataset as more individuals are turning to reduce their carbon footprint by purchasing electric vehicles.

### 2. Related Work

With an ever-growing push from many cities to pull their weight in the global decarbonization effort. Shifting over to Electric vehicles and reducing greenhouse gas emissions (GHG) as well as petroleum emissions has become a focal point for many large cities like New York. Urban areas currently account for 70% of GHG emissions globally, so reducing their emissions would play a large role in the effort to fight global warming. It should come as no surprise that many organizations, cities, and interested parties are conducting various research to understand what cities are doing to address this in terms of programming.

The first study we found that uses the Electric Vehicle Drive Clean Rebate from the New York State Energy Research and Development Authority (NYSERDA) data for their analysis is titled "Disparities and equity issues in electric vehicles rebate allocation" by Shuocheng Guo and Eleftheria Kontou [1]. Their study focuses on the availability and magnitude of monetary incentives, such as rebates, and how they positively impact PEVs market penetration, but most importantly, it focused on equitable policies that could be implemented to ensure that income and disparities would be accounted for to ensure that rebates are not just distributed in more affluent and less disadvantaged census tracts.

The second study is an extended benefit-cost analysis done on the deployment of electric vehicles in the state of New York using a similar dataset to ours. This study included models on EV adoption, costs of electric vehicles, fueling costs, charging rates and fees, the investments to be made in charging infrastructures, and emission factors for key pollutants like GHG emissions. The study included recent survey information that reflects some of our concluded analysis. For instance, a recent AAA survey indicated that 20% of Americans are interested in purchasing an electric vehicle [2]. The Fuels Institute also reported in 2017, that more than 50% of potential car buyers said they were very or somewhat likely to purchase an all-electric vehicle. Our analysis conclusion in Figure 3, Lease and Buy Analysis vs. Rebate amount per make, reflects that the majority of electric vehicles are purchased over leasing. This recent benefit-cost analysis updated many of the modeling assumptions, methodology, and key inputs compared to work previously done in New York State. The study benefited from almost five years of additional market data like electric vehicle consumer purchasing habits, consumer charging behavior, and updated fuel pricing, including electricity and gasoline. The larger the dataset, the higher the accuracy of deep learning can achieve. Our data set was much smaller; therefore, limited features and conclusions could be drawn.

### 3. Specifications

The dataset was retrieved from Data.gov [3], a U.S. Government website that aims to improve public access to high-value, machine-readable datasets generated by the Federal Government's Executive Branch. The site contains information available to the public from federal, state, local, and tribal governments. The dataset was provided by NYSERDA, which offers objective information and analysis, innovative programs, technical expertise, and support to help New Yorkers increase energy efficiency, save money, use renewable energy, accelerate economic growth, and reduce reliance on fossil fuels. Dealers enrolled in the program deduct the eligible amount from the vehicle price at the point of sale and then submit a rebate application with NYSERA. This dataset includes all completed rebate applications as of 2017 up to date. The dataset consists of makes, models, counties, zip codes, electric vehicle types, transaction types, annual GHG emissions reductions, annual petroleum reductions in gallons and rebate amounts in US dollars of the different electric vehicles. The size of the dataset is 6.91MB.

### 4. Implementation Flowchart

The raw dataset downloaded from NYSERDA comprised of total rebate amount, annual petroleum emission, and annual greenhouse gas emissions per county and zip code for New York state. The data was retrieved from Data.gov. The process of data manipulation is shown in the flowchart below. The dataset being used was provided in a single CSV file. The data file was uploaded to the SAP Analytics cloud which was used to clean and create a story with models. The models were then exported to be used in PowerPoint.

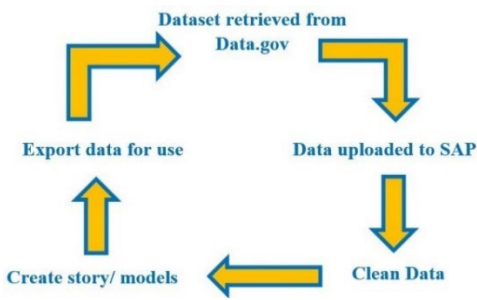

Figure 1. Implementation Flowchart

### 5. Data Cleaning

The CSV file was uploaded onto SAP Analytics cloud as a dataset. From there, the first step is to transform the original dataset to include additional columns that allow future data analysis and visualization. First, a "State" column was added to indicate all the data are from New York State. Next, a "Count" column was created, assigning a numeric value of 1 for each data entry. This column will be used to summarize the number of rebate submissions under various conditions. Then, the hierarchy relationship of geological locations was established under **Action** -> **Geo Enrich** -> **Area Names**. Finally, the whole dataset was reviewed, and all entries were retained for use in creating stories in SAP.

### 6. Analysis and Visualization

After data cleaning and determining what data would best describe our analysis, a story, and predictive model
were created in SAP Analytics Cloud that provided a visual representation of the car make, model that participate in the rebate program, the types of electric vehicle and purchase preference by different car make, the environmental impact of electric vehicle, and the financial preparation on the government side to meet the growing demand of electric vehicle from consumers.

### 6.1. Total Rebate Amount (USD) Per Make

First, we wanted to know which vehicle maker would benefit the most from this rebate program. To find the answer to our question, we created a pie chart as shown in Figure 2,. This pie chart was created in SAP Analytics Cloud and shows the difference in rebate amounts depending on the vehicle make.

Figure 2 shows that the highest rebate amount is reserved for vehicles make Tesla while the lowest rebate amount was given to BMW (based on make received at least 1.6% of the total rebate). In reviewing the chart, we can see that the top 5 vehicle makes that received the highest volume of rebates are the following in order from largest to smallest: Tesla (42.49%), Toyota (20.07%), Chevrolet (11.15%), Hyundai (8.75%) and Ford (4.56%).

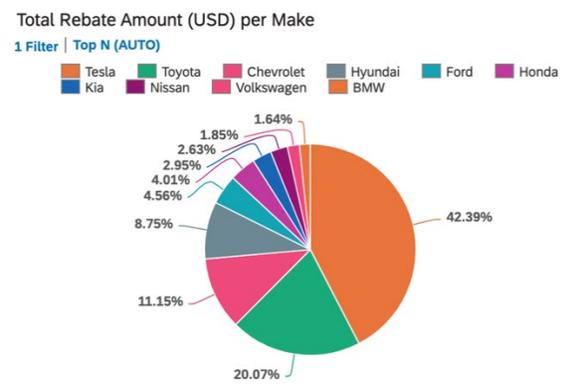

Figure 2. Total Rebate Amount (USD) per Make

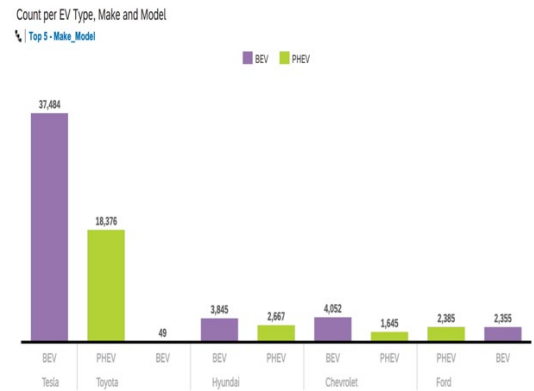

Figure 3. County per EV Type, Make and Model

### 6.2. Count per EV Type, Make & Model

An EV is defined as a vehicle that can be powered by an electric motor that draws electricity from a battery and is capable of being charged from and make Toyota showed a different profile. The popular electric vehicle type from Toyota is PHEV. Even Toyota still offers BEV, and it is not a popular choice among consumers. For other top makes, such as Hyundai, Chevrolet, and Ford, both BEV and PHEV have occupied a significant percentage within the same make.

### 6.3. Lease and Buy Analysis vs. Rebate Amount per Make

Our first two questions and findings shared some very insightful information about the top-performing make and rebate amount given per make. It led us to want to learn more about how many total registered EVs are on the road, not just EVs but the total number of EVs per maker, and whether New Yorkers were purchasing or leasing the vehicles. Additionally, we wanted to see the comparison to the Rebate amount.

To answer this question, we used the NYSERDA Electric Vehicle Drive Clean Rebate Data in Figure 4 which illustrates the difference in rebate amount in US dollars per make and whether these vehicles were purchased or leased. The chart illustrates and shares a familiar finding that Tesla, the top maker in our last chart, received rebate amounts mainly due to purchased vehicles. In a similar pattern, Toyota was the second leading purchased vehicle, followed by Chevrolet. Interestingly, in contrast to the top 3 makers, Hyundai users received more rebate dollars from leases than purchases. It demonstrates the consumer's behavior differences when choosing different electric vehicle brands.

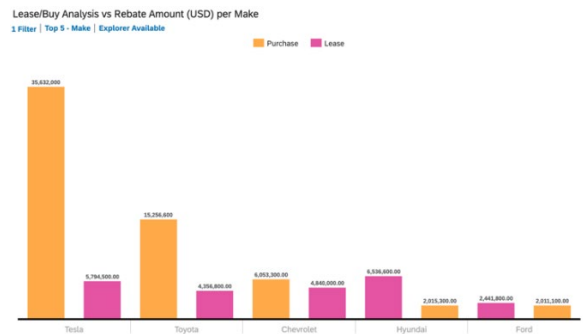

Figure 4. Lease/Buy Analysis vs. Rebate Amount (USD) per Make4

### 6.4. Annual GHG Emissions (MT CO2e) and Annual Petroleum Reductions (gallons) and others per Make and Model

Following our initial question, we wanted to find out which vehicle is shown to reduce the most GHG (MT CO2e) and Petroleum (gallons) emissions. In Figure 4, we evaluate all the makers and group them into different clusters. The Cluster Map highlights the vehicle makes with the highest annual GHG emissions reductions and petroleum reductions in gallons. Among all the makes, Tesla distinguishes itself from the rest as the one with the highest reduction in annual GHG emission (MT CO2e) and petroleum reduction (group 3, mint color) external source. BEVs are fully electric vehicles that rely on battery power and use no gasoline. While PHEVs use battery packs for driving shorter distances, often around 30 miles. When the battery depletes, then the gas-powered engine takes over. After finding out total rebates per make, annual greenhouse and petroleum emission reductions per maker and model, and county. We would like to understand total PEV and PHEV sales to understand EV adoption rates per make and model. Figure 5 shows the count per EV type, maker, and model for this. We can see that Tesla is still leading like it has throughout our analysis. It is leading with BEV, which are fully electric vehicles that rely on battery power and no gasoline. However, the second, the make coming after Tesla, is Toyota, which also forms group 2 in pink. The rest of the makers form group 1 in orange due to similar environmental impact.

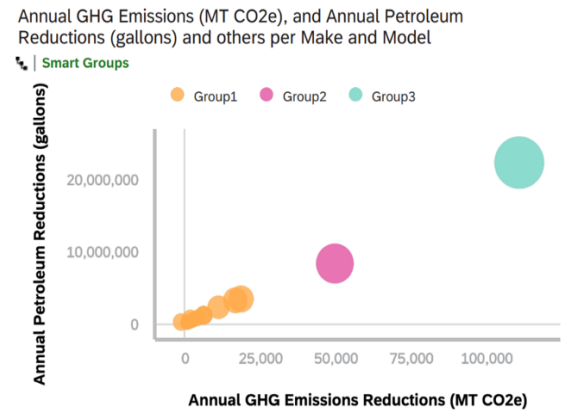

Figure 5. Annual GHG Emissions Reductions (MT CO2e), Annual Petroleum Reductions (gallons) and others per Model for Actual

### 6.5. Annual GHG Emissions Reductions (MT CO2e) by Area (subregion/county)

We found out the top makers of vehicles that reduce environmental externalities for all and the incentives, such as electric vehicle rebates, where they are distributed, and in what amounts. Now, we want to learn (1) how these numbers and findings were distributed, and (2) where and what county was showing to reduce their GHG emissions the most.

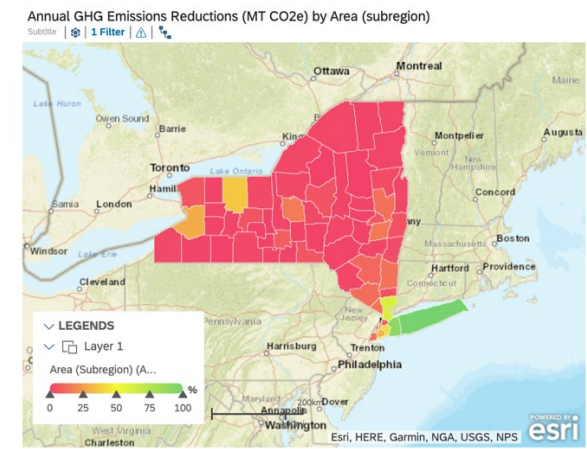

Figure 6. Annual GHG Emissions Reductions (MT CO2e) by Area (subregion)

It can provide more insight into other questions about equity distribution amongst allocation between income groups and disadvantaged communities for rebates and how it impacts counties in New York differently. Figure 6 clearly shows the subregions/ counties with the lowest GHG emissions reduction (MT CO2e) and the Highest GHG reduction in a 4-color gradience. This data shows that Suffolk has the highest GHG drop at 36,779.06 (MT CO2e), followed by Westchester with the second highest GHG reduction at 22.567.14 (MT CO2e). The map in Figure 6 shows that most New York State does not reduce GHG emissions as much.

### 6.6. Time Series Analysis

State-level rebate incentives reduce financial barriers between prospective owners and increase EV sales for BEVs and PHEVs. To facilitate the adoption of electric vehicles, the government should allocate sufficient funding for the rebate program. Thus, we illustrate an optimistic upward trend in rebate amounts since the inception of the program in Figure 7. It shows the amount of rebate issued by the government based on the submission date. In addition, the forecast function and triple exponential smoothing were used to predict the amount of funding needed for the year 2023 on a quarterly basis.

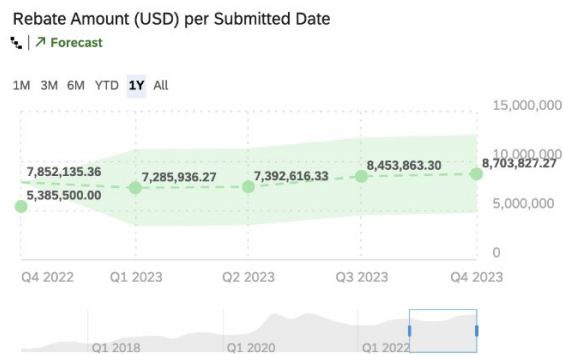

Figure 7. Rebate Amount (USD) per Submitted Date

### 7. Future work

The dataset we used in this analysis is relatively small, with only over eighty-eight thousand entries. With more information from other regions, we should also confidently predict the sales of Electric Vehicles in the future. In addition, this dataset can be combined with other information, such as the MSRP of vehicles, charging station density, and population data, to analyze further the factors that impact consumers' decision-making in purchasing electric vehicles.

### 8. Conclusion

In reviewing and summarizing all the above charts, maps and analyses conducted we conclude the following:
I. The top 5 vehicle makes with the greatest rebate amounts (from largest to smallest) are:
- Tesla, with a rebate of $41,426,500
- Toyota, with a rebate of $19,613,400
- Chevrolet, with a rebate of $10,893,300
- Hyundai, with a rebate of $8,551,900
- Ford, with a rebate of $4,452,900

II. The top two (2) counties of the state of New York with the highest reductions of GHG emissions due to electric vehicles are Suffolk at 36,779,06 (MT CO2e) and Westchester at 22,567.14 (MT CO2e). However, the rest of the state is not reducing GHG emissions enough.
III. Tesla, followed by Toyota, is the electric vehicle proven to reduce the most GHG (MT CO2e) and petroleum (in gallons) emissions.
IV. Rebate amounts per participant have continued to rise since the inception of the rebate program.
V. The highest purchased electric vehicle, versus leasing, is Tesla, followed by Toyota.
VI. Tesla leads in battery electric vehicles by count compared to other vehicle makes.